\documentclass[useAMS,usenatbib]{mnras}
\usepackage{graphicx,lscape,amssymb}
\usepackage{natbib}
\usepackage{fancyhdr}
\usepackage{subfigure}
\usepackage{longtable}
\usepackage{rotating}
\usepackage{calc}
\usepackage{fancyhdr}
\usepackage{float}
\usepackage{enumerate}
\usepackage{soul}
\usepackage{sidecap}

\usepackage{amsmath}
\usepackage{subfigure}
\usepackage{booktabs, dcolumn}
\usepackage[dvipsnames]{xcolor}
\usepackage{hyperref}

\newcommand{\Teff}{\mbox{$T_{\mathrm{eff}}$}} 
\newcommand{\Msun}{\mbox{$\mathrm{M_{\odot}}$}}

\usepackage{soul}

\newcommand{\bp}{\mbox{$G_{\rm BP}$}}
\newcommand{\rp}{\mbox{$G_{\rm RP}$}}

\title[White dwarfs as IR standards]{Cool white dwarfs as standards for infrared observations}

\author[Gentile Fusillo et al.]  
{Nicola Pietro Gentile Fusillo,$^{1}$\thanks{E-mail: N.Gentile-Fusillo@warwick.ac.uk} 
Pier-Emmanuel Tremblay,$^{1}$ 
Ralph C. Bohlin,$^{2}$ 
\newauthor 
Susana E. Deustua,$^{2}$ 
and Jason S. Kalirai,$^{3}$ 
\newauthor 
  \\
  $^{1}$Department of Physics, University of Warwick, CV4 7AL, Coventry, UK\\
  $^{2}$Space Telescope Science Institute, 3700 San Martin Drive, Baltimore, MD 21218, USA\\
  $^{3}$Johns Hopkins Applied Physics Laboratory, 11100 Johns Hopkins Road, Laurel, MD 20723, USA\\}
  
\begin{document}
\maketitle

\label{firstpage}
\begin{abstract}
In the era of modern digital sky surveys,  uncertainties in the flux of stellar standards are commonly the dominant systematic error in photometric calibration and can often affect the results of higher-level experiments. The Hubble Space Telescope (HST) spectrophotometry, which is based on computed model atmospheres for three hot (\Teff\ $> 30\,000$\,K) pure-hydrogen (DA) white dwarfs, is currently considered the most reliable and internally consistent flux calibration. 
However many next generation facilities (e.g. Harmoni on E-ELT, Euclid and JWST) will focus on IR observations, a regime in which white dwarf calibration has not yet been robustly tested. 
Cool DA white dwarfs have energy distributions that peak close to the optical or near-IR, do not have shortcomings from  UV metal line blanketing, and have a reasonably large sky density ($\simeq 4$\, deg$^{-2}$ at  $G<20$), making them, potentially, excellent calibrators. 
Here we present a pilot study based on STIS+WFC3 observations of two bright DA white dwarfs to test whether targets cooler than current hot primary standards (\Teff\ $<20\,000$\,K) are consistent with the HST flux scale. We also test the robustness of white dwarf models in the IR regime from an X-shooter analysis of Paschen lines and by cross-matching our previously derived Gaia white dwarf catalogue with observations obtained with 2MASS, UKIDSS, VHS, and WISE.

\end{abstract}
\begin{keywords}
white dwarfs - stars: fundamental parameters - infrared: general - line: profiles - stars: individual: WD1327$-$083,  stars: individual: WD2341+322
\end{keywords}
\section{Introduction}

In the last 20 years the advent of large area digital surveys revolutionised the way we observe the sky, and today astronomers can rely on terabytes of imaging data covering almost the entire sky from the ultraviolet (UV) to the infrared (IR). Modern electronic detectors allow a precision of $<1$ per cent in the determination of the physical energy distributions of stars and surveys like SDSS, PanSTARRS and Gaia  can boast  internal  relative  photometric  accuracies  on the scale of 1-2 per cent \citep{padmanabhan08-1, panstarrs16-1,gaiaDR2-ArXiV-3}. However any attempt to compare observations across different surveys and in particular across different wavelength coverage ultimately requires for all observations to be placed on a single reliable and consistent flux scale.
The uncertainties in the flux of such absolute stellar standards are often the dominant systematic error in photometric calibration and propagate into the higher-level analysis  of  several experiments.   

For example our understanding of the nature of the dark energy, the driving force behind the observed accelerating cosmic expansion, fundamentally relies on the accurate comparison between the fluxes of type Ia supernovae at different redshifts and therefore observed at different wavelengths \citep{riessetal98-1, perlmutter99-1,riessetal00-1}. Quantitative descriptions of dark energy in terms of the Einstein equations of general relativity are significantly improved when the relative flux with wavelength is known to an accuracy of 0.2 per cent  or better \citep{scolnicetal14-1}. Currently, the most precise and internally consistent set of fluxes are the Hubble Space Telescope (HST) spectrophotometry, primarily from STIS and NICMOS, which are based on computed model atmospheres for
three hot (\Teff\ $> 30\,000$~K) hydrogen atmosphere (DA) white dwarfs, GD71, GD153, and G191-B2B
\citep{bohlinetal14-1}. The absolute flux at 5556 \AA~is fixed by ground-based measurements
adjusted slightly to match MSX IR absolute fluxes \citep{bohlin14-1}, but the relative flux at
other wavelengths is entirely determined by the three hot white dwarf standards. Recent studies have largely confirmed the internal consistency of such calibration procedure with fainter secondary hot white dwarf standards \citep{narayanetal19-1}. Thus, our understanding of cosmology and the nature of the dark energy is tied to the precision of model spectral energy distributions (SEDs) of hot white dwarfs. 

There is a re-assuring internal consistency of $\simeq1$ per cent between the predicted hot white dwarf SEDs and the HST flux scale in the optical range (3500-8500 \AA). However other optical photometric surveys such as Gaia, PanSTARRS, and SDSS do not directly rely on the HST flux scale, resulting in a miss-match of white dwarf parameters when relying on different fitting techniques \citep{tremblayetal19-1}. Independent verifications of white dwarf fluxes are thus still needed to confirm the accuracy of the HST flux scale and improve the calibration of external photometric surveys. Such task is challenging and in this work we focus in particular on cooler and redder hydrogen-atmosphere white dwarfs. This aims at enhancing the accuracy of the HST flux scale and providing insight on the near- and mid-IR calibration regimes that have not been robustly tested with white dwarfs. 

Many of next generation facilities which will see first light in the next decade will focus on IR observations. The first-generation instrument HARMONI on E-ELT will be the telescope's workhorse for spectroscopy covering the wavelength range 0.47-2.45 $\mu$m \citep{harmoni16-1}. The Euclid spacecraft will survey 15\,000 deg$^2$ of the sky targeting hundreds of thousands white dwarfs as faint as 24\,mag and the NISP instrument will provide near-IR photometry in $Y, J, H$ bands, and low-resolution spectroscopy between 0.92 and 1.85  $\mu$m \citep{euclid16-1}. The array of instruments on the James Webb Space Telescope (JWST) will carry out photometric and spectroscopic observations both in the near- and mid-IR regime down to 30 mag allowing to target the faintest and coolest white dwarfs even at the edge of the Milky Way \citep{jwst18-1}.
Additionally, uncertainties in white dwarf fluxes are currently dominant in using the Wide Field InfraRed Survey Telescope (WFIRST) to precisely constrain the nature of dark energy (see figure 9 of \citealt{hounselletal18-1}). It is therefore of paramount importance to define precise flux standards in this wavelength regime.
Hot white dwarfs have a low sky density ($\simeq$ 0.75\,deg$^{-2}$ at $G<20$) and their model atmospheres suffer from uncertainties due to non-local thermal equilibrium
(NLTE) effects and the presence of metal traces and UV line blanketing \citep{gianninasetal10-1}. Consequently, HST's spectrophotometric white dwarf standards may not be ideal comparison objects for IR observations; and other types of white dwarfs or stars could be used as near-IR flux calibrators \citep{bohlinetal14-1}. However, recent WFC3 near-IR observations of hot white dwarfs are in good agreement with the HST flux scale \citep{narayanetal19-1}.

Following upon the preliminary analysis of \citet{tremblayetal17-1}, here we test whether DA white dwarfs cooler than current hot primary standards (\Teff\ $ < 20\,000$\,K) are consistent with the HST flux scale. Cool H-atmosphere white dwarfs have largely featureless energy
distributions that peak in the optical or near-IR, and their colors are more similar to those of high-redshift supernovae or galaxies. Cool white dwarfs do not have shortcomings from NLTE effects or UV metal line blanketing, have a reasonably large sky density ($\simeq$ 4\,deg$^{-2}$ at $G<20$), and are generally closer than their hot counterparts, thus resulting in less or negligible reddening. In contrast, as white dwarfs cool below \Teff\ $\approx 13\,000$\,K they develop convective atmospheres which affect the measurement of their stellar parameters. State-of-the-art grids of 3D model atmospheres can now reliably account for convective effects \citep{tremblayetal13-1} as verified by Gaia \citep{tremblayetal19-1}.
Cool H-atmosphere stellar remnants have, therefore, a strong potential as IR calibrators. We base our study on STIS+WFC3 and X-Shooter observations of a small sample of bright DA white dwarfs. We also review the empirical and predicted near- and mid-IR white dwarf Hertzprung-Russell diagram by cross-matching the extensive Gaia white dwarf catalogue of \citet{gentilefusilloetal19-1} with the Two Micron All Sky Survey (2MASS), the UKIRT Infrared Deep Sky Survey (UKIDSS), the Vista Hemisphere Survey (VHS), and the Wide-field Infrared Survey Explorer (WISE). This provides further insights on the accuracy of white dwarf models at these wavelengths.

\section{Hubble Space Telescope spectrophotometry}

\subsection{STIS and WFC3 observations}

In cycle 23 we obtained 4 orbits of STIS and WFC3 observations (GO-14213) of two intermediate temperature (12\,500\,K $\lesssim$ \Teff\ $\lesssim$ 15\,000\,K) H-atmosphere white dwarfs: WD1327$-$083 and WD2341+322. We selected objects that were bright enough to achieve a 1 per cent flux precision similar to current hot standards \citep{bohlinetal14-1} in two orbits. We have excluded ZZ Ceti pulsators (10\,500\,K $\lesssim$ \Teff\ $\lesssim$ 12\,500\,K at $\log g$ = 8.0 ($\log g/$[cm s$^{-2}$]), see \citealt{tremblayetal15-2}) and suspected binaries. We also exclude magnetic white dwarfs as these objects are known to show photometric and spectroscopic variability and have SEDs which are extremely challenging to model (e.g. \citealt{brinkworthetal13-1}). Finally, white dwarfs with trace metals (DAZ) were also discarded as possible targets. A small fraction of these stars are known to exhibit IR emission from a dusty debris disc \citep{becklinetal05-1} which would make the target unsuitable for IR calibration.
The selected targets are in the intermediate temperature regime between current hot white dwarf standards and cooler non-pulsating H-atmospheres (\Teff\ $<$ 10\,500\,K). These objects are cool enough to have negligible NLTE effects and seemingly no metal opacity in the UV, yet they are not in the regime where 3D corrections due to convection are significant \citep{tremblayetal13-1}. In the future we hope to secure similar HST observations for even cooler bright white dwarfs and in the meantime we will explore the calibration potential of cooler white dwarfs with existing photometric observations in Section~\ref{gaia}.

STIS spectroscopy of our two stars with the G140L, G230L, G430L, and G750L gratings covers the range 1150--10000~\AA\ at a resolution of R = 500--1000. Our
STIS spectrophotometry is augmented by WRC3 IR grism observations with coverage from 8000 to 17000~\AA\ using G102 and G141 with resolutions R = 200 and 150, respectively \citep{bohlinetal19-1}. The STIS data are processed with the updated calibration procedure of Bohlin, Deustua, and de Rosa (2019 in prep.), while details of the WFC3 grism calibration procedure appear in  \citet{bohlinetal15-1} and in \citet{bohlinetal19-1}. The STIS and WFC3 SEDs are merged at 10120 and 9700~\AA\ for WD1327$-$083 and WD2341+322, respectively, to make complete HST SEDs for 
CALSPEC\footnote{http://www.stsci.edu/hst/instrumentation/reference-data-for-calibration-and-tools/astronomical-catalogs/calspec}.

\subsection{Analysis}

Setting up DA white dwarfs as flux standards generally implies first determining \Teff, surface gravity (expressed as $\log g$ where $g$ is in units of cm/s$^{2}$),  radius and reddening in order to define model absolute fluxes. The spectroscopic technique of fitting the observed Balmer lines with model SEDs \citep{bergeronetal92-1} has been widely regarded as the most precise method to determine \Teff\ and surface gravity of white dwarfs, which can then be converted to radii using the well understood white dwarf mass-radius relation. Relative uncertainties in these measurements are typically of the order of $<2$ per cent for high signal-to-noise observations (S/N $\gtrsim 40$; \citealt{liebertetal05-1}) and are largely unaffected by reddening. The true accuracy of the atmospheric parameters is difficult to measure as it depends intimately on the accuracy of the line broadening physics \citep{tremblayetal09-1,gomez16}. However absolute fluxes predicted by spectroscopy are in agreement with observed dereddened Gaia fluxes at a few per cent level \citep{tremblayetal19-1,genestetal19-1}. 

In contrast white dwarf \Teff\ and radius can be determined by photometric analyses using broadband colours and trigonometric parallaxes \citep{koesteretal79-1,bergeronetal01-1}. The advantage of this technique is that model fluxes for a given \Teff\ and radius are fairly robust because they depend on better known continuum opacities rather than line broadening physics. Furthermore, the recent Gaia DR2 provided a large sample of precise photometric and parallax measurements for white dwarfs allowing for relative \Teff\ and radius uncertainties that are now at the same precision level as spectroscopic analyses \citep{gentilefusilloetal19-1}. The drawback  of this technique is that photometric measurements are subject to reddening and must be correctly calibrated to obtain reliable results. Currently, calibration issues appear to be the dominant uncertainty in using photometric solutions \citep{tremblayetal19-1,bergeronetal19-1}. Photometric calibration is often performed by comparing with white dwarf parameters obtained by fitting spectroscopy \citep[see, e.g.,][]{holberg+bergeron06-1}, however this approach does not necessarily allow for better accuracy in the atmospheric parameters.
Here we use HST STIS+WFC3 spectrophotometry of two intermediate \Teff\ white dwarfs to compare different methods of estimating atmospheric parameters.

Our observations have been calibrated based on the current three hot white dwarf standards (\Teff\ $>30\,000$\,K) in CALSPEC. Our targets have significantly lower \Teff\ values, hence have energy distributions that peak closer to the optical regime and that are not affected by continuum hydrogen opacities in the same way as hot white dwarfs. Therefore they can serve as a test of the accuracy of the CALSPEC flux scale and possibly define on their own new primary standards. 

We fitted the newly acquired STIS+WFC3 spectra of  WD1327$-$083 and WD2341+322 using our grid of H-atmosphere models \citep{tremblayetal11-1,tremblayetal13-1}. The model atmospheres include Lyman, Balmer, Paschen and Brackett line-blanketing (40 lines in total) and use up to 1814 frequencies to solve radiative transfer for the atmospheric structures in the Rosseland optical depth range $-5\lesssim \log \tau_{\rm R} \lesssim 3$. We use the non-ideal equation-of-state of \citet{hummer+mihalas88-1} both for populations and opacities \citep{tremblayetal09-1}. The grid spacing is given in section 2.2 of \citet{tremblayetal11-1}

We follow the standard spectroscopic fitting method by comparing continuum-normalized Balmer line profiles in the white dwarf spectra with our synthetic spectral models. Additionally, since STIS and WFC3 are spectrophotometrically calibrated, we also compare the entire STIS spectrum directly with our models without any normalization step. In this second procedure we rely on the Gaia DR2 \citep{gaiaDR2-ArXiV-1,gentilefusilloetal19-1} parallax of the stars to scale the synthetic spectrum and directly fit the STIS flux to obtain a radius estimate. We then use the white dwarf mass-radius relationship \citep{fontaineetal01-1} to derive masses ($0.61\pm0.01$\,\Msun\ and $0.62\pm0.02$\,\Msun\ for WD1327$-$083 and WD2341+322 respectively) and log $g$ values for the stars and compare with the spectroscopic solutions. The models which best match the entire SED of the two white dwarfs are provided with the electronic distribution of this article. As shown in Table\,\ref{param} the \Teff\ and log $g$ values obtained using both methods are in excellent agreement ($<1\sigma$). The fact that these two fitting routines produce results in such good agreement indicates that the STIS+WFC3 spectra are not affected by interstellar reddening. Indeed no reddening should be expected for stars as close as WD1327$-$083 and WD2341+322 (16.1\,pc and 18.6\,pc respectively). 

Stellar parameters for these white dwarfs have also previously been published in \citet{gianninasetal11-1} using independent spectroscopic observations and in \citet{gentilefusilloetal19-1} using Gaia observations only.  For  WD1327$-$083 we find that our best solution closely agrees ($<1 \sigma$) with both previously published values. For  WD2341+322 our model analysis of the  STIS spectrum results in parameters which are in excellent agreement with those from \citet{gentilefusilloetal19-1}, but are 2$\sigma$ discrepant from the \citet{gianninasetal11-1} solution (Table\,\ref{param}). 

\begin{table*}
\centering
\caption{\label{param} Previously published and newly calculated atmospheric parameters of the white dwarfs WD1327$-$083 and WD2341+322. All quoted uncertainties correspond to $1\sigma$.}
\begin{tabular}{ll D{?}{\,\pm\,}{5.3} D{?}{\,\pm\,}{5.3}  c D{?}{\,\pm\,}{5.3} D{?}{\,\pm\,}{5.3}}
\hline
& Method & \multicolumn{2}{c}{\textbf{WD1327$-$083}} & & \multicolumn{2}{c}{\textbf{WD2341+322}}\\
& & \multicolumn{1}{c}{\Teff\ [K]} & \multicolumn{1}{c}{log $g$} & &  \multicolumn{1}{c}{\Teff\ [K]} & \multicolumn{1}{c}{log $g$}\\

\hline \\[-1.5ex]
\multicolumn{2}{l }{\citet[][Balmer lines]{gianninasetal11-1}} & 14\,570?240 & 7.99?0.04 & & 13\,100?198 & 7.92?0.04\\
\multicolumn{2}{l }{Gaia parallax + photometry \citep{gentilefusilloetal19-1}}& 14861?107 & 7.99?0.01 & & 12\,650?43 & 8.01?0.05\\
\multicolumn{2}{l }{STIS+WFC3 Balmer lines (Figs.\,\ref{WD1327-083_full}, \ref{WD2341+322_full})} & 14\,837?406 & 8.01?0.05 & & 12\,884?214 & 8.06?0.04\\
\multicolumn{2}{l }{STIS+WFC3 continuum + Gaia parallax fit (Figs.\,\ref{WD1327-083_full}, \ref{WD2341+322_full})} & 14\,607?290 & 7.99?0.03 & & 12\,671?330 & 8.02?0.03\\
\multicolumn{2}{l }{STIS+WFC3 Paschen lines (Fig.\,\ref{WF3_paschen})} & 14822?830 & 8.00?0.1 & & 13\,805?840 & 8.05?0.1\\

\hline\\
\end{tabular}
\end{table*}

\begin{figure*}
\includegraphics[width=1.9\columnwidth]{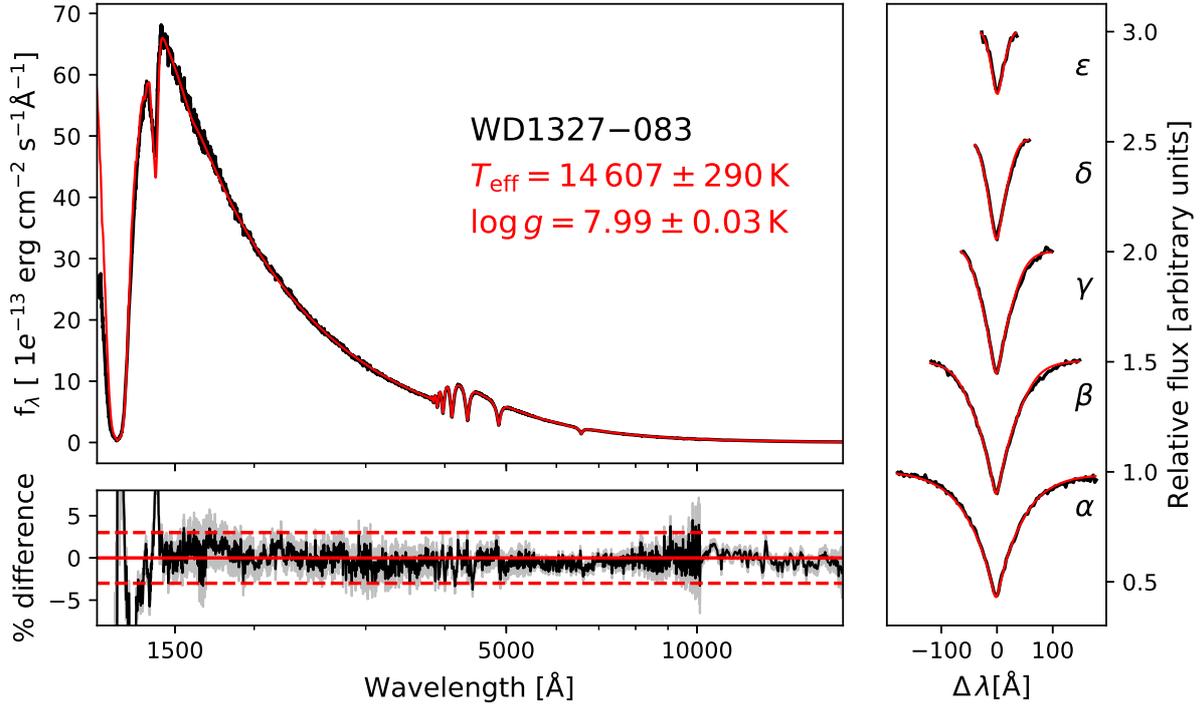}
\caption{\label{WD1327-083_full} \textit{Main panel}: STIS+WFC3 spectrophotometry of the white dwarf WD1327$-$083 with the best fitting model and best fitting parameters displayed in red. \textit{Bottom panel}: Residuals in the difference between the observed flux and the model predicted flux. Uncertainties in the STIS+WF3 observed flux are represented by the gray-shaded region. The red dashed lines indicate the 3 per cent level. \textit{Right panel}: Best model fit to the Balmer lines. The line profiles are vertically offset for clarity. The best fit parameters are found in Table\,\ref{param}.}
\end{figure*}

\begin{figure*}
\includegraphics[width=1.9\columnwidth]{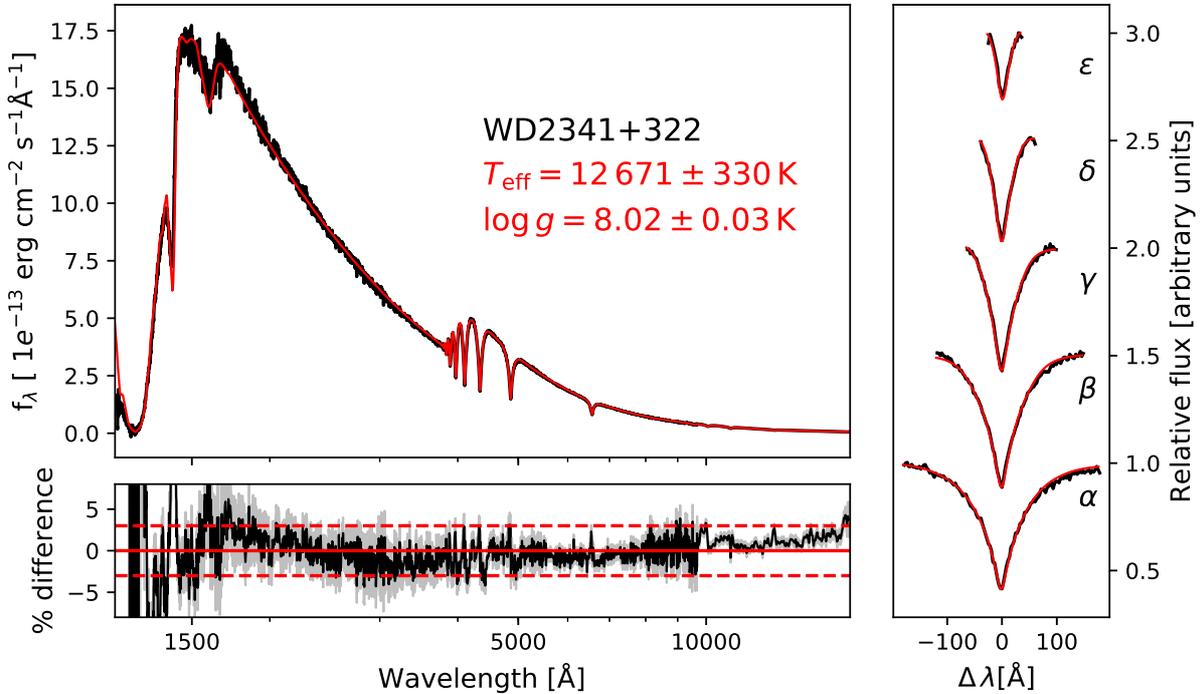}
\caption{\label{WD2341+322_full}
Similar to Fig.\,\ref{WD1327-083_full} but for STIS+WFC3 spectrophotometry of the white dwarf  WD2341+322. }
\end{figure*}

We have used our own grid of model atmospheres, but to our knowledge other grids of H-atmosphere white dwarfs use the same constitutive physics in this temperature range. \citet{tremblayetal11-1} show that their code fully connects with TLUSTY \citep{hubeny+lanz95-1} which is used at higher temperatures, and \citet{tremblayetal17-1} have verified that the model atmosphere code of \citet{koester10-1}  provides similar spectroscopic and photometric solutions within $1 \sigma$ for WD\,1327$-$083. The systematic uncertainties from subsets of the input physics (e.g. line profiles) may be larger than the difference between the codes, but that is difficult to quantify.

Figs.\,\ref{WD1327-083_full} and \ref{WD2341+322_full} clearly illustrate that, overall, our models successfully predict the observed STIS and WFC3 fluxes to within 3 per cent over most of the range between 1500\,\AA\ and 17\,000\,\AA. In particular in the near-IR between 8000\,\AA\ and 16\,000\,\AA\ the model and observed fluxes diverge by less than 2 per cent for both white dwarfs. Overall our results suggest that there is no fundamental reason to limit white dwarf calibrators to \Teff\ $>$ 20\,000~K \citep{narayanetal19-1}. Using a lower temperature limit would represent a five-fold enhancement in the sky density of white dwarf calibrators.

\section{Paschen series fitting}

The so-called spectroscopic method, which relies on comparing observed Balmer line profiles with model spectra, is widely regarded as a reliable technique to obtain \Teff\ and log $g$ for H-atmosphere white dwarfs. In the near-IR  between  8500 and 13\,500~\AA\ the spectra of white dwarfs display absorption features from the lower-energy Paschen series. These line profiles can also be compared with model spectra and could in principle have the same diagnostic potential as Balmer lines. Testing whether current white dwarf model atmospheres correctly replicate observed Paschen line profiles and whether stellar parameters obtained by fitting these lines are compatible with spectroscopic fits using Balmer lines has two purposes. First of all the capability of white dwarfs as infrared calibrators rests on the ability of the model atmospheres to fit the observed data. Secondly in the next decade key instruments onboard facilities such as JWST and Euclid will have IR-only capabilities, and Paschen lines could offer unique diagnostic opportunities for observed H-atmosphere white dwarfs.

According to the \citet{hummer+mihalas88-1} formalism, interactions between neighboring particles cause atomic energy levels to have a certain probability of being dissociated, and in such case the absorption of a photon  results in a bound-free transition (so-called pseudo-continuum opacity) instead of a regular line transition. In warm H-atmosphere white dwarfs where Paschen lines are observed, these non-ideal effects are caused by the interactions of the H atoms with charged protons and electrons. As a consequence the higher Lyman, Balmer and Paschen lines can be much weaker than expected from an ideal equation-of-state. \citet{tremblayetal09-1} were the first to combine Stark broadening with a consistent implementation of line dissolution according to the \citet{hummer+mihalas88-1} model. The implementation of these improved Stark profiles has been extensively discussed for the Balmer lines \citep{tremblayetal09-1} and the Lyman lines \citep{prevaletal15-1}. Here we include for the first time Paschen line profiles calculated under the formalism of \citet{tremblayetal09-1}. While these models have not been used by our group before, we note that broadband near-IR photometry is not impacted by the details of line broadening, and therefore previously available colour tables are unchanged.

We developed a fitting routine for the Paschen lines of H-atmosphere white dwarfs. In order to construct a sample of test objects we obtained VLT X-shooter \citep{vernetetal11-1} observations of 10 bright H-atmosphere white dwarfs. The spectra were acquired in period 97 between April and July 2016  utilizing nodding mode, a $1''$ slit aperture for the UVB arm and $0.9''$ for the VIS arm and $1.2''$ for the NIR arm. Exposure times ranged between 190\, and 1190\,s. The spectra were reduced using the standard procedures within the {\sc reflex}\footnote{http://www.eso.org/sci/software/reflex/} reduction tool developed by ESO \citep{freudlingetal13-1}.
The combined UVB, VIS and NIR arms of X-shooter cover the entire spectral range from $\simeq300$0 to $14\,000$ \AA\  which allows fitting both the Balmer and Paschen sequences. The X-shooter targets spanned a broad range of \Teff, but our observations revealed that at $\Teff \lesssim 8000$ K and $\Teff \gtrsim 40\,000$ K the Paschen absorption lines become too weak for any meaningful model comparison.  Because of this restriction we were able to compare stellar parameters only for 8 of our 10 stars (Table\,\ref{X-fits}).
\begin{table*}
\centering
\caption{\label{X-fits} Atmospheric parameters of the white dwarfs observed with X-shooter on the VLT. All quoted uncertainties correspond to 1\,$\sigma$.}
\begin{tabular}{l D{?}{\,\pm\,}{5.3} D{?}{\,\pm\,}{5.3} c  D{?}{\,\pm\,}{5.3} D{?}{\,\pm\,}{5.3}}
\hline
 Name & \multicolumn{2}{c}{\textbf{Balmer lines fit}} & & \multicolumn{2}{c}{\textbf{Paschen lines fit}}\\
& \multicolumn{1}{c}{\Teff\ [K]} & \multicolumn{1}{c}{log $g$} &  & \multicolumn{1}{c}{\Teff\ [K]} & \multicolumn{1}{c}{log $g$}\\
\hline \\[-1.5ex]
WD1229$-$012&20380?385&7.57?0.05& &19450?610&7.51?0.1\\
WD1241$-$010*&24696?265&7.41?0.06& &24561?430&7.51?0.1\\
WD1244$-$125&13254?357&7.97?0.04& &12990?800&8.01?0.06\\
WD1310$-$305&20315?317&7.91?0.05& &19590?610&7.90?0.07\\
WD1356$-$233&9644?150&8.27?0.07& &9818?180&8.32?0.1\\
WD1407$-$475&22668?448&7.78?0.06& &25360?730&7.69?0.1\\
WD1418$-$088&8090?153&8.29?0.04& &8178 ?160&8.36?0.07\\
WD1500$-$170&32591?526&7.94?0.05& &35000?990&7.73?0.1\\ 
\hline\\
[-1.5ex]\multicolumn{4}{l}{\small{*Double white dwarf binary \citep{marshetal95-1}}}\\
\end{tabular}
\end{table*}

\begin{figure*}
\includegraphics[width=1.8\columnwidth]{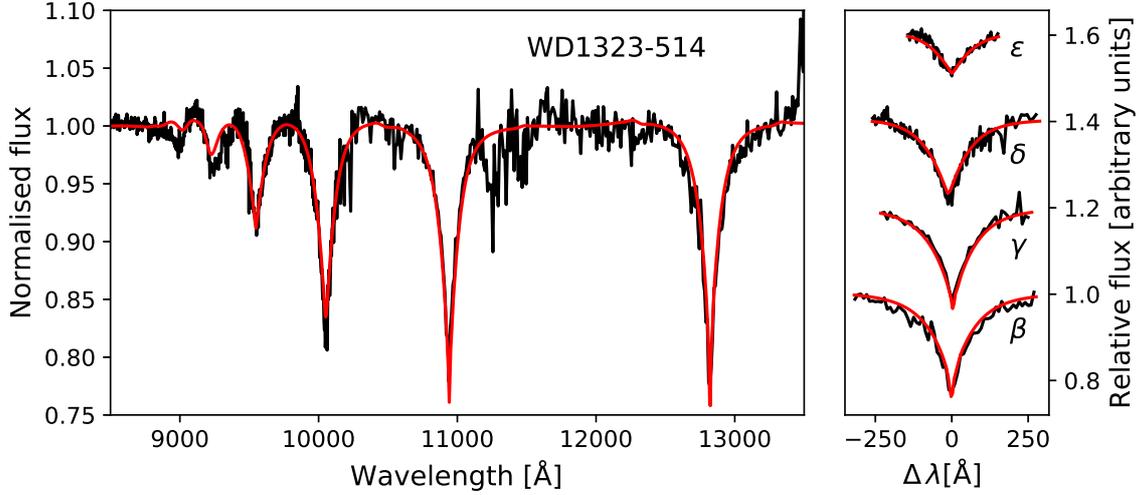}
\caption{\label{Xs_paschen} Atmospheric model fit to the Paschen lines from the telluric corrected, normalised, X-shooter spectrum of the white dwarf WD1323$-$514 (left panel). Residuals from telluric line removal can still be seen between 12\,000 and 13\,000 \AA. 
The individual lines used for the fit are displayed in the right panel, vertically offset for clarity.}
\end{figure*}

Analogously to the well-established Balmer spectroscopic fitting method, our newly developed Paschen fitting routine requires to first normalize the continuum of the spectrum of the white dwarf and then compare the individual Paschen lines to  model spectroscopy (Fig.\,\ref{Xs_paschen}). Ground-based spectroscopic observations of the Paschen lines suffer from heavy contamination by atmospheric absorption features. Telluric lines removal is therefore crucial in order to proceed with any model comparison. We used \textsc{MOLECFIT} \citep{molecfit15-1} 
 to remove tellurics, but despite the overall success in cleaning the spectra, the results are not uniform across all objects and some residuals still affect the spectroscopy of some our targets. Residuals in the telluric line subtraction can affect the shape of the Paschen lines as well as the continuum normalization step resulting in less accurate stellar parameters.
In our model comparison we only rely on the Pa$\epsilon$, Pa$\delta$, Pa\,$\gamma$ an Pa$\beta$ lines. Though visible in some the spectra, the higher order lines are too weak for a meaningful comparison with the models and Pa$\alpha$ is completely hidden by atmospheric features and cannot be recovered. 

\begin{figure*}
\includegraphics[width=1.8\columnwidth]{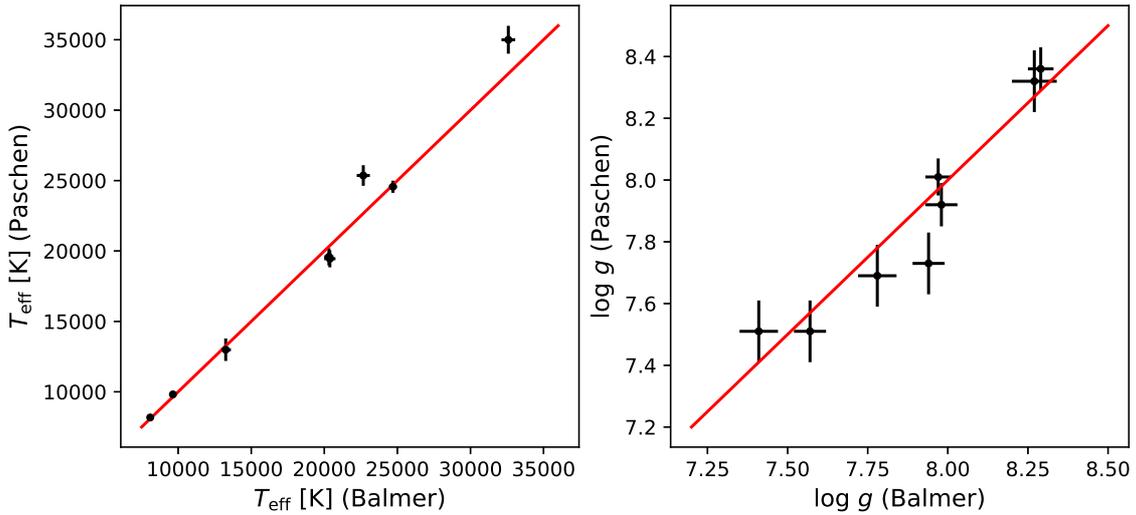}
\caption{\label{NIRvsVIS} Comparison between \Teff\ (left panel) and log $g$ (right panel) obtained by fitting Balmer and Paschen lines for 8 bright H-atmosphere white dwarfs observed with X-shooter (see Table\,\ref{X-fits}). The red lines represent the 1:1 relationship.}
\end{figure*}

\begin{figure*}
\includegraphics[width=1.8\columnwidth]{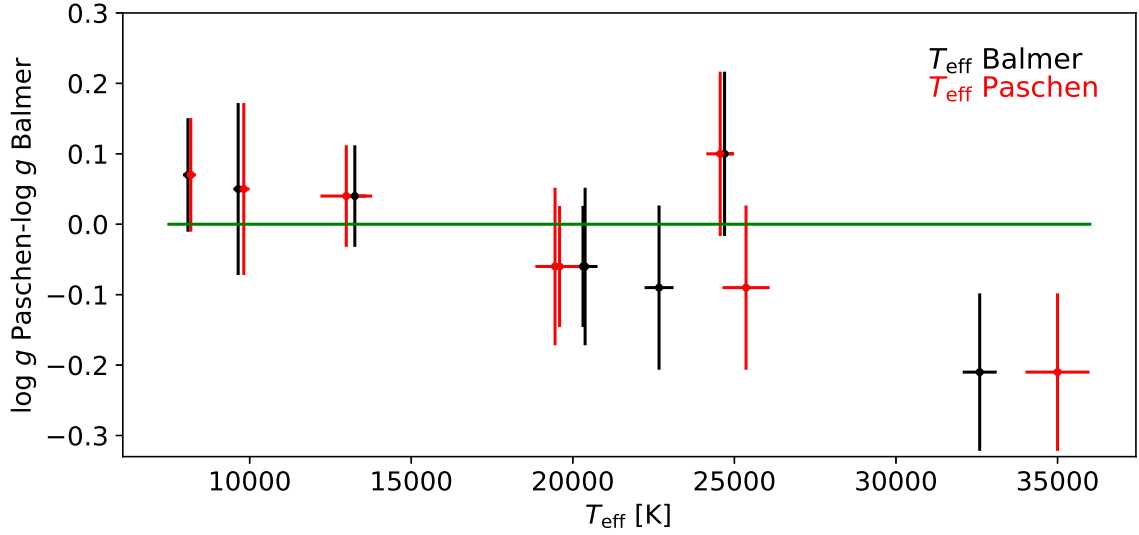}
\caption{\label{ref_test} Comparison of the log\,$g$ values obtained by fitting Paschen lines and Balmer lines, as a function of \Teff. Each object is included at the position of its \Teff value from Balmer lines fit (black) and from Paschen lines fit (red). The green horizontal line indicates the relationship  log\,$g$ (Paschen)=log\,$g$ (Balmer).}
\end{figure*}

\begin{figure*}
\includegraphics[width=1.9\columnwidth]{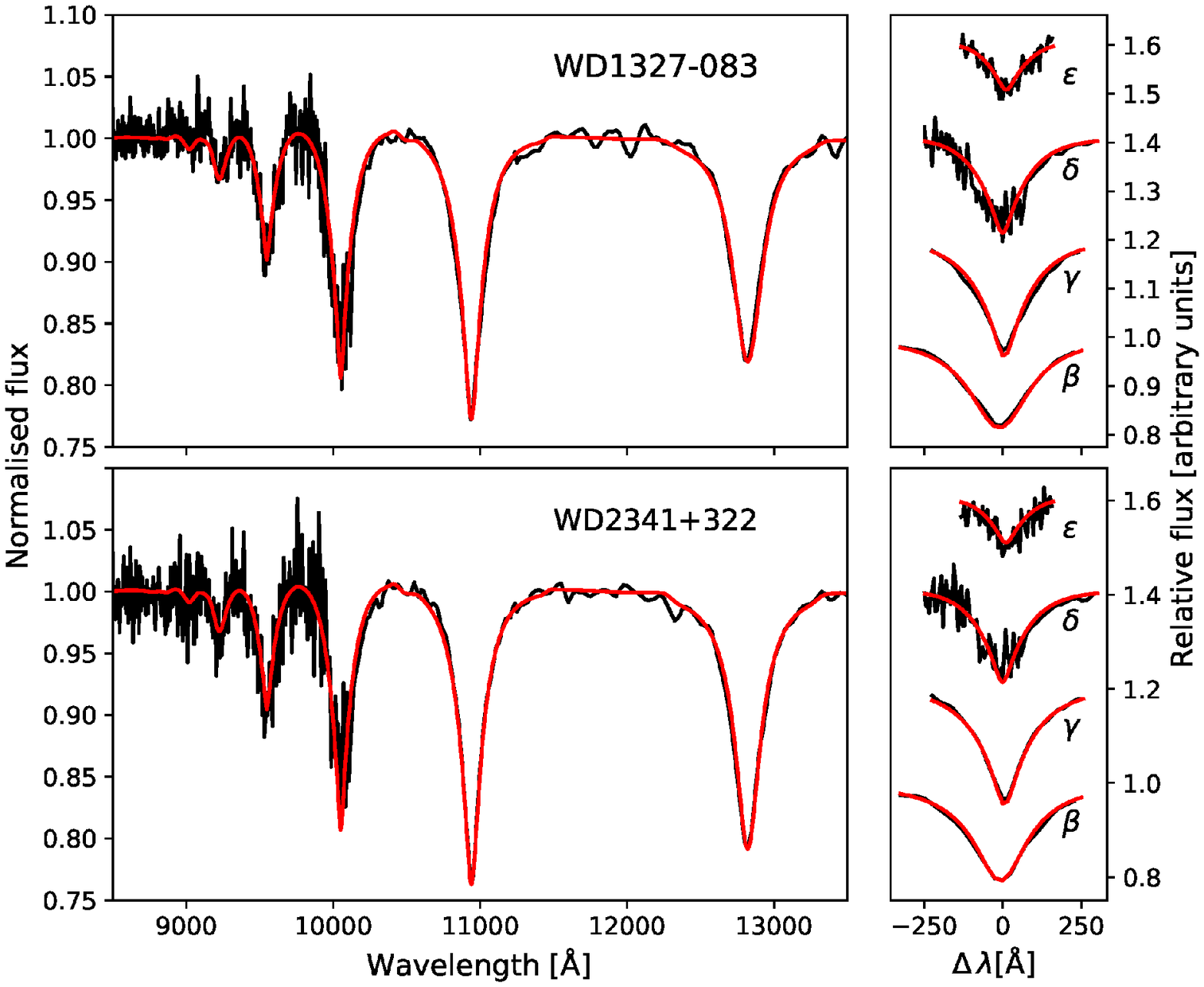}
\caption{\label{WF3_paschen} Atmospheric model fit to the Paschen lines from the STIS+WFC3 spectra of the white dwarfs WD1327$-$083 and WD2341+322. The individual lines used for the fit are displayed on the right panels, vertically offset for clarity.}
\end{figure*}

Fig.\,\ref{NIRvsVIS} shows that we find a good overall agreement between the stellar parameters obtained from the Paschen and Balmer line fits, but some solutions are significantly discrepant ($> 3 \sigma $). The outlying objects have particularly strong telluric absorption features in their X-shooter spectra and the subtraction procedure could not fully account for them.  We observe a possible trend indicating that, with increasing \Teff, Paschen lines fits return lower log\,$g$ values compared to Balmer lines fits (Fig\,\ref{ref_test}). However, with only 8 white dwarfs in the sample this trend is potentially only apparent because of the outlier nature of WD1500$-$170. This is the hottest white dwarf examined and so the object with the weakest Paschen lines in the sample, making it particularly susceptible to imperfect telluric removal.
Nonetheless this comparison illustrates that our models correctly predict the line profiles of Paschen lines within uncertainties and near-IR spectroscopy can be used to estimate the stellar parameters of H-atmosphere white dwarfs. Space based spectroscopy would be best suited for this type of observations as it would be free from the limitations introduced by the presence of telluric lines.
The combined STIS+WFC3 spectra of  WD1327$-$083 and WD2341+322 show clear Paschen absorption lines and so provide an additional opportunity to further test the reliability of our white dwarf models. 
As shown in Table\,\ref{param} the stellar parameters obtained from the Paschen fits (Fig.\,\ref{WF3_paschen}) are in good agreement with those obtained from the Balmer lines and continuum fits. However it important to point out that due to the low resolution of the WFC3 near-IR spectra the uncertainties in stellar parameters obtained from the Paschen lines are relatively high. 


\section{Ground-based Near-IR and WISE mid-IR photometry}
\label{gaia}
Near-IR photometry can also be employed to test the accuracy of white dwarf models in this wavelength regime. While the calibration of ground-based photometric surveys may not be as well as understood as HST STIS and WFC3, the large number of observed white dwarfs, in many instances with the same object observed by two or more surveys, provides an advantage compared to the smaller HST data set.
To date a number ground-based surveys provide near-IR photometric coverage in the $Y, J, H ,K$ bands, for vast areas of the sky. In this work we focus on 2MASS, UKIDSS, and VHS as they provide the coverage of the largest number of white dwarfs. 

2MASS covers the entire sky to a limiting magnitude of 15.8, 15.1 and 14.3 in $J, H$ and $K_{s}$, respectively. Nearly 20 years after its nominal end 2MASS remains a very powerful tool as the only all-sky near-IR survey \citep{skrutskieetal06-1}. Nonetheless, the relatively shallow magnitude depth limits the survey usefulness for intrinsically faint objects like white dwarfs.

UKIDSS is the collective name of a group of five near-IR surveys: the Large Area Survey (LAS), the Galactic Plane Survey (GPS), the Galactic Cluster Survey (GCS), the Deep Extragalactic Survey (DXS) and the Ultra Deep Survey (UDS). Because of its large footprint UKIDSS LAS is best suited for our work. It provides imaging over 4000 deg$^2$ north of Dec = $-$1.25 deg in four broad bands: $Y, J, H$ and $K$, with limiting magnitudes of 20.5, 20.0, 18.8 and 18.4, respectively \citep{lawrenceetal07-1}. 

VHS is the widest area near-IR survey conducted from the VISTA telescope at Paranal (Chile). It scans the sky in the $Y, J, H$ and $K_{s}$ bands \citep{VHS13-1} and the most recent data release (DR6) covers almost the entire southern hemisphere. 

Space based observations in the mid-IR regime offer the opportunity to test white dwarf models at even longer wavelengths, but resources for large scale comparisons are much scarcer compared to the near-IR.
The only reliable large-area survey in the mid-IR is WISE, which mapped the entire sky  at 3.4, 4.6, 12 and 22 $\mathrm{\mu m}$ in four bands ($W1, W2, W3$ and $W4$; \citealt{wrightetal10-1}). However most white dwarfs only have reliable detection in the $W1$ band and the low spatial resolution of WISE ($\simeq 6''$) makes large scale cross-matching particularly challenging. 

\begin{figure*}
\includegraphics[width=1.8\columnwidth]{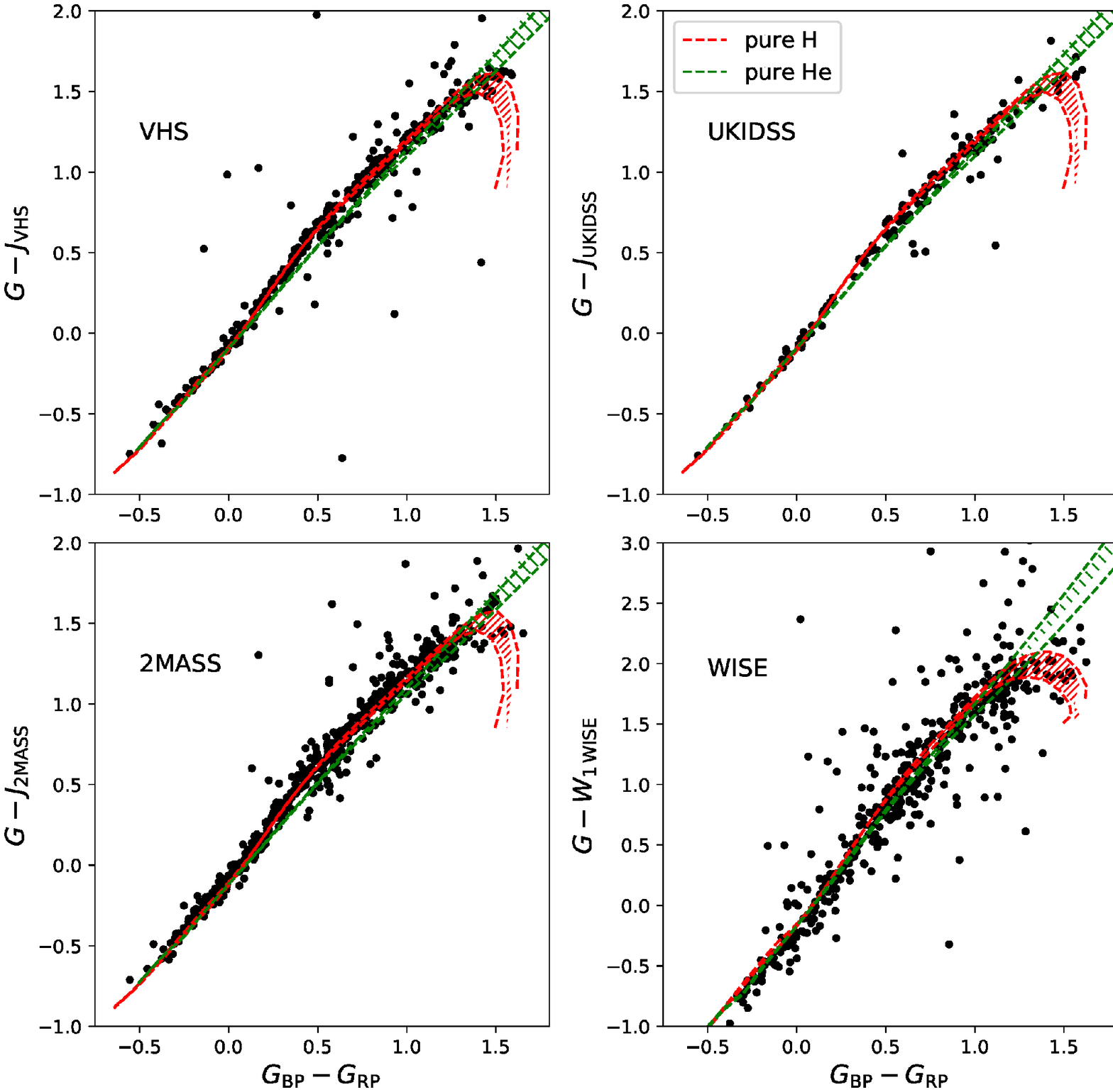}
\caption{\label{col_col} Colour-colour diagrams combining Gaia $G$, $G_{\rm BP}$ and $G_{\rm RP}$ magnitudes with $J$-band measurements from VHS, UKIDSS and 2MASS, and with $W1$-band from WISE for white dwarfs within 50\,pc of the Sun observed by each survey. Cooling tracks for pure H and pure He atmosphere white dwarfs are displayed in red and green respectively. The distribution of white dwarfs in these colour-colour spaces is mostly insensitive to $\log g$. The shaded regions between the cooling tracks show the area spanned by objects with $7.5<\log g<8.5$. The predicted colours are in good agreement with the observed ones for all surveys.}
\end{figure*}

In order to compare IR fluxes predicted by current models with actual IR observations, we cross-matched the volume-limited 50\,pc sample of high confidence white dwarf candidates ($P_{\rm WD} \geq $ 0.75)  from the \citet{gentilefusilloetal19-1} catalogue with the aforementioned four IR surveys. 
Given the relatively large wavelength difference between the Gaia $G$ band and near-IR $J$ bands, the comparison of the empirical and modelled $G-J$ colours can be a powerful tool to confirm the consistency of predicted model fluxes. The broad wavelength separation also means that $G-J$, $G-H$ and $G-K$ colours have very similar features but with decreasing precision, so for illustrative purposes we focus on $G-J$ (and $G-W1$ for WISE).
The colour-colour diagrams in Fig.\,\ref{col_col}, using Gaia \bp $-$ \rp~ colour as an independent variable, clearly illustrate that the synthetic colours correctly match the observed white dwarf sequence in all four surveys. The larger scatter seen for WISE is caused by the relatively worse photometric data quality and potentially by a small number of mis-matchings. The only significant difference between the photometric data and the models is the ``hook" due to collision induced absorption in the pure hydrogen atmosphere tracks which appear not to be populated by any white dwarf. The most likely cause is that the very cool (\Teff\ $<4000$\,K) white dwarfs which would occupy this area of the diagram are rare due to the finite age of Galactic disc, cool rapidly, and in most cases are too faint to be detected in the IR surveys. We note that such colour-colour diagrams have little sensitivity to surface gravity and even atmospheric composition as shown in Fig.\,\ref{col_col}, in clear contrast with the Hertzprung-Russell diagrams discussed in the next section.

\subsection{IR Hertzprung-Russell diagrams}
\begin{figure*}
\includegraphics[width=1.8\columnwidth]{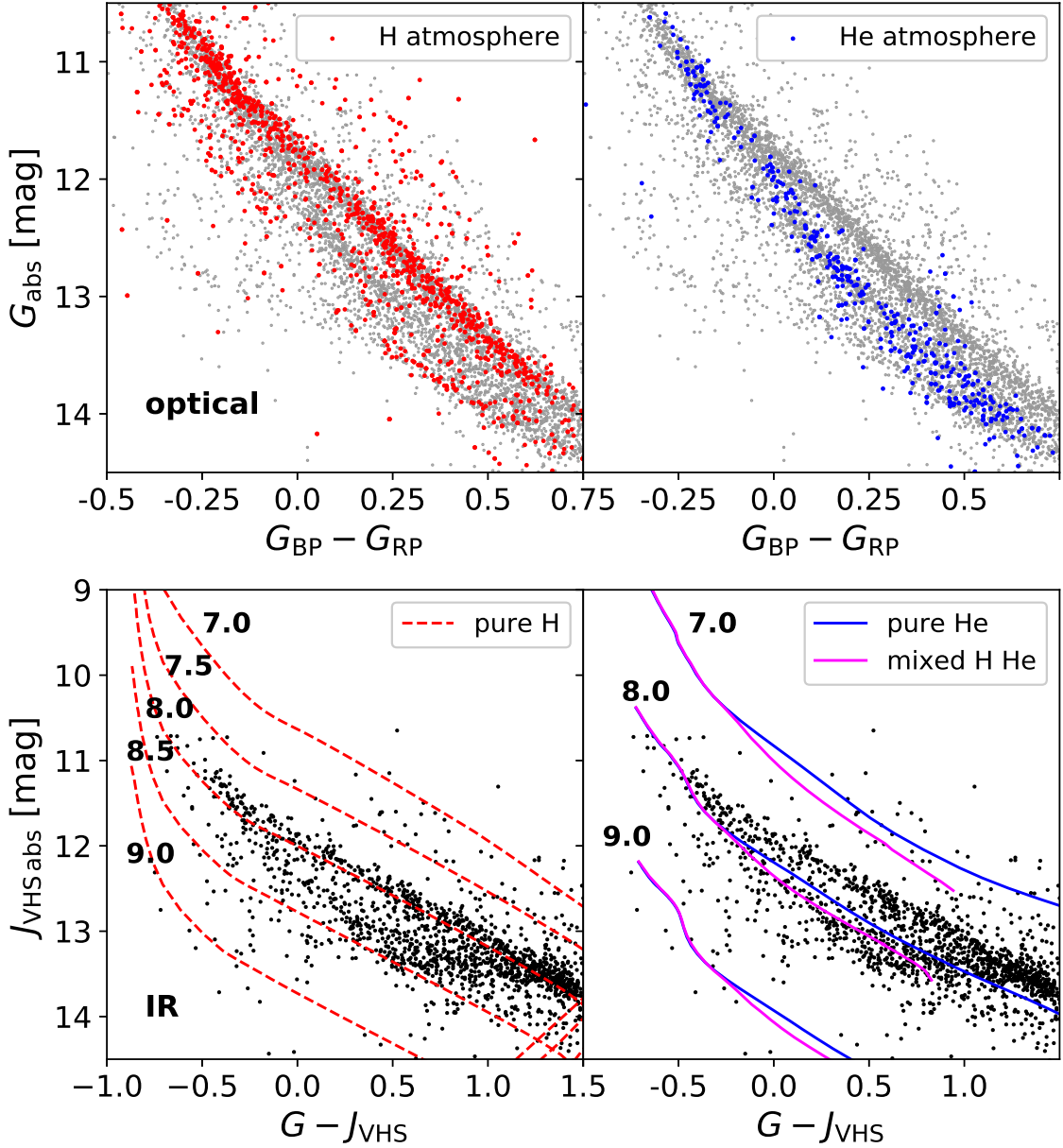}
\caption{\label{HR_J} 
\textit{Top panels:}  H-R diagrams showing the distribution of Gaia white dwarfs using optical colours. High-confidence white dwarf candidates within 50\,pc of the Sun from \citet{gentilefusilloetal19-1} are displayed in grey. To highlight the bifurcation between white dwarfs with different atmospheric composition, pure H and He-atmosphere white dwarfs identified using Sloan Digital Sky Survey (SDSS) spectroscopy are shown in red (top left) and blue (top right) respectively. 
\newline \textit{Bottom panels:} Distribution in the H-R diagram of white dwarfs within 50\,pc of the Sun with available VHS $J$-band photometry. The bifurcation between pure H and He-rich atmosphere white dwarfs is clearly visible also in IR-colours for $G-J_{\rm VHS} \gtrsim -0.25$. \textit{Bottom left panel}: cooling tracks for pure H atmosphere white dwarfs at different $\log g$ values are displayed in red showing a good agreement between the predicted $\log g \approx 8.0$ value and observed IR fluxes for DA white dwarfs in the upper luminous branch of the bifurcation. \textit{Right panel}: cooling tracks for pure He and mixed (H/He = 10$^{-5}$ by number) atmosphere white dwarfs at different $\log g$ values are displayed in blue and magenta respectively. Pure He models (blue) at a $\log g \approx 8.0$ value clearly fail to reproduce the observed IR fluxes in the lower branch of the bifurcation while the mixed models (magenta) achieve a better agreement. Mixed atmosphere tracks have been truncated at \Teff\ $= 5000$\,K, below which they strongly differ from the observations. }
\end{figure*}

 The top panels in  Fig.\,\ref{HR_J} clearly illustrate the bifurcation, in the optical-colours Hertzprung-Russell diagram, between DA white dwarfs of typical surface gravity (top sequence) and the equivalent track of He-atmosphere white dwarfs (bottom sequence). Combining Gaia measurements with IR photometry also allows to inspect the H-R diagram distribution of white dwarfs in IR colour space. The bifurcation between DA white dwarfs and He-atmosphere white dwarfs is clearly visible even at IR-wavelengths (Fig.\,\ref{HR_J}). Analogously to what we showed for the optical H-R diagram in \citet{gentilefusilloetal19-1}, Fig.\,\ref{HR_J} also illustrates that, while the DA cooling track can be modelled well, pure-He model atmospheres do not correctly reproduce the cooling sequence of the He-atmosphere lower branch. Mixed model atmospheres with H/He $= 10^{-5}$ in number result in a better agreement with the observed IR data in Fig.\,\ref{HR_J}, similarly to what was demonstrated for the optical H-R diagram in \citet{bergeronetal19-1}. 

Below \Teff\ = 5000\,K the effect of collision induced absorption predicted by our mixed models is far too strong leading to a marked discrepancy with the observed data. More recent models described in  \citet{blouinetal19-1} appear to have solved this low temperature discrepancy by including non-ideal ionisation. A better comparison with IR photometry of low temperature He-atmosphere white dwarfs may be possible once the predicted colours of these new model grids become available. Nonetheless, the need to include additional degrees of freedom (H abundance and/or metals), when modelling He-atmosphere white dwarfs, makes them more challenging to characterize, and so less suitable as calibrators than the more common DA stars.

\section{Conclusion}
We have conducted a pilot test of the reliability of intermediate temperature and cool white dwarfs (\Teff\ $<20\,000$\,K) as calibrators, specifically in the IR regime. We have analysed newly acquired STIS+WFC3 HST spectrophotometry of the two bright white dwarfs WD1327$-$083 and WD2341+322 and find that current atmosphere and interior models correctly predict the observed fluxes of both objects within 3 per cent over the entire wavelength range from 1500 to 17\,000\,\AA. We therefore conclude that intermediate temperature white dwarfs are fully consistent with the HST flux scale and, in most cases, could be used as calibrators that are equally reliable as the current hot primary standards. The synthetic spectra which best match the STIS+WFC3 SED of WD1327$-$083 and WD2341+322 are available with the electronic distribution of this article.  
The much higher space density of cooler white dwarfs, $\simeq 4$\, deg$^{-2}$ compared to a value of $\simeq 0.75$\, deg$^{-2}$ for their hot counterparts, represents a significant advantage when looking for suitable calibrators for both large surveys and single observations. Additionally, since the spectral energy distribution of cool white dwarfs peaks much closer to the near-IR compared to the hot flux standards, these objects will be better suited as calibrators for the observations conducted by the next generation IR facilities like Euclid and JWST.

To reliably test the robustness of white dwarf models in the IR regime we ran two additional experiments: a model fit to the Paschen line profiles and a comparison with IR photometric observations.
For the Paschen line fits we have developed a technique analogous to the well established Balmer line spectroscopic method and vetted its reliability using a small sample of H-atmosphere white dwarfs observed with the X-shooter spectrograph on the VLT. We have then fitted the Paschen line profiles to obtain stellar parameters that only rely on near-IR spectroscopic data. For all stars we find \Teff\ and $\log g$ values in close agreement with those obtained from the  analysis of their optical and UV spectra. This opens up a new venue of analysis for white dwarfs in the near-IR regime. For the second test we have compared the optical versus IR colour-colour diagrams computed from our white dwarf models with the observed distribution of these objects within 50\,pc as observed by Gaia, 2MASS, UKIDSS, VHS and WISE. We show that the model tracks closely match the observations for all surveys. While the implication of these results depends on our understanding of the calibration of these surveys, this suggests that our models adequately predict IR fluxes of cool white dwarfs.

\section*{Acknowledgements}
The research leading to these results has received funding from the European Research Council under the European Union's Horizon 2020 research and innovation programme n. 677706 (WD3D).  IR190/1-1.
This work has made use of data from the European Space Agency (ESA) mission {\it Gaia} (\url{https://www.cosmos.esa.int/gaia}), processed by the {\it Gaia} Data Processing and Analysis Consortium (DPAC, \url{https://www.cosmos.esa.int/web/gaia/dpac/consortium}). Funding for the DPAC has been provided by national institutions, in particular the institutions participating in the {\it Gaia} Multilateral Agreement. 
The work presented in this article made large use of TOPCAT and STILTS Table/VOTable Processing Software \citep{topcat-1}. We thank the anonymous referee for his constructive review. 

\appendix
\section{Observation log}
\label{obs_log}
\begin{table*}
\centering
\caption{\label{param} Log of HST observations}
\begin{tabular}{ccccc}
\hline
Object & Telescope/Instrument &grating & exposure time [s] & Date [dd-mm-yyyy] \\
\hline \\[-1.5ex]

\textbf{WD1327$-$083} &HST/STIS &G750L& 936.0& 15-04-2016\\
 &HST/STIS &G430L& 200.0& 15-04-2016\\
 &HST/STIS &G230L& 605.0& 15-04-2016\\
 &HST/STIS &G140L& 1244.0& 15-04-2016\\
&HST/WFC3 &G141& 60.14& 15-04-2016\\
 &HST/WFC3 &G102& 83.51& 15-04-2016\\
\hline\\[-1.5ex]
\textbf{WD2341+322}&HST/STIS &G750L& 939.0& 27-05-2016\\
 &HST/STIS &G430L& 240.0& 27-05-2016\\
 &HST/STIS &G230L& 650.0& 27-05-2016\\
 &HST/STIS &G140L& 1370.0& 27-05-2016\\
 &HST/WFC3 &G141&128.45 &27-05-2016\\
 &HST/WFC3 &G102&175.98& 27-05-2016\\

\hline\\
\end{tabular}
\end{table*}

\bibliographystyle{mnras}
\bibliography{aamnem99,aabib1,aabib2,aabib_new,aabib_tremblay}
\end{document}